\documentclass[a4paper,12pt,preprint,nofootinbib]{revtex4}
\usepackage{amsmath,mathrsfs,amscd,graphicx}
\usepackage{mciteplus}
\usepackage{multirow}
\usepackage{color}
\usepackage{rotating}
\usepackage{slashed}
\usepackage{subfig}
\usepackage{textcomp}
\usepackage[final]{pdfpages}
\usepackage{array}
\usepackage{float}
\usepackage{url}
\usepackage{textcomp}
\usepackage{amsfonts, latexsym, epsfig}
\usepackage{bm}
\usepackage{times}
\usepackage{epsfig}
\usepackage{amssymb}
\usepackage{tikz}
\usepackage{cancel}
\usepackage{hyperref}
\usepackage{verbatim}
\usepackage[normalem]{ulem}
\usepackage{diagbox}

\def\beq{\begin{equation}}
\def\eeq{\end{equation}}
\def\be{\begin{equation}}
\def\ee{\end{equation}}
\def\bea{\begin{eqnarray}}
\def\eea{\end{eqnarray}}

\begin{document}

\title{Searching for the light Higgsinos {in MSSM} at the future e-p colliders}
\author{Chengcheng Han~$^{a,b}$}
\email{chengcheng.han@ipmu.jp}
\author{Ruibo Li~$^{a}$}
\email{bobli@zju.edu.cn}
\author{Ren-Qi Pan~$^{a}$}
\email{renqipan@zju.edu.cn}
\author{Kai Wang~$^{a}$}
\email{wangkai1@zju.edu.cn}
\affiliation{
$^a$ Zhejiang Institute of Modern Physics, Department of Physics, Zhejiang University, Hangzhou, Zhejiang 310027, CHINA\\
$^b$ Kavli IPMU (WPI), The University of Tokyo, Kashiwa, Chiba 277-8583, JAPAN
}
\preprint{IPMU18-0029} 

\begin{abstract}
The search of the light Higgsino {in the Minimal Supersymmetric Standard Model (MSSM)} is a crucial test for the criteria of the naturalness in supersymmetry. On the other hand, the direct production of light Higgsino is also known as one of the most challenging SUSY searches at the current CERN Large Hadron collider (LHC). The lack of visible leptons due to the compressed spectrum and their small production rates limits their discovery potential in both mono-jet plus $\cancel{E}_{T}$ as well as the weak boson fusion (WBF) production. The signal $S/B$ ratio prediction is usually within the background systematic uncertainties at the LHC. Without color exchange between the beams, the $e-p$ colliders are well-known to have the WBF feature. Therefore, we study the search of the light Higgsinos  at two future $e-p$ colliders at CERN, LHeC and FCC-eh.~The light Higgsinos will be produced in pair through weak boson fusion with controlled background at these colliders. We find the Higgsino of 95/145~GeV can be reached at 2$\sigma$ level at the future LHeC/FCC-eh with a luminosity  $3 ~\text{ab}^{-1}$ respectively. 
\end{abstract}

\maketitle
\section{\label{intro}Introduction: light Higgsino survived from direct searches}

The ATLAS and CMS experiments at the CERN Large Hadron Collider (LHC) have put stringent lower bounds for the supersymmetric (SUSY) particles which excluded large parameter space for the low energy SUSY models. 
As a consequence of conserved $R$-parity, SUSY particles are usually produced in pair through gauge interactions at the LHC.~Searches of final state with jets or leptons plus $\cancel{E}_{T}$ have excluded the ${\cal O}$(1.5 -- 2~TeV) colored SUSY particles production at this QCD machine~\cite{Aaboud:2017bac,Aaboud:2017vwy}.~The direct production of electroweak SUSY particles has relatively small cross section but the definite hard lepton tagging also helps the search in the multi-lepton plus $\cancel{E}_{T}$ final states and ${\cal O}$(600~GeV) charginos has been excluded \cite{Aaboud:2017nhr}. One well-known exception is the light Higgsinos with nearly-degenerate spectrum and relatively small production rate\footnote{Light Higgsinos may reside in the theory from the naturalness argument \cite{Arnowitt:1992qp}.}.


In the limit of $\mu \ll M_1, M_2$,  the low energy chargino/neutralinos are all Higgsino-like and nearly degenerate. If $M_1, M_2 >$ 1.5 TeV and  $\mu \sim$ 100 GeV, the mass-splitting, $m_{\tilde{\chi}^0_2}-m_{\tilde{\chi}^0_1}$ is less than 5 GeV while the mass splitting of $m_{\tilde{\chi}^\pm_1} - m_{\tilde{\chi}^0_1}$ is about half of that. The key property for these nearly degenerate electroweak-inos 
 of $\Delta m = m_{\tilde{\chi}^\pm_1} - m_{\tilde{\chi}^0_1}\sim {\cal O}$(GeV) is the 
lack of hard leptons in the final states. Therefore, the final states of $\tilde{\chi}^{\pm}_{1}\tilde{\chi}^{0}_{2}$, $\tilde{\chi}^{\pm}_{1}\tilde{\chi}^{0}_{1}$ and $\tilde{\chi}^{+}_{1}\tilde{\chi}^{-}_{1}$, etc.~mostly manifest themselves as $\cancel{E}_{T}$ \footnote{If the $\Delta m > 5$~GeV, soft muons sometimes can serve as an additional handle to suppress the background \cite{Giudice:2010wb,Han:2014kaa,Baer:2014kya,Han:2015lma}. }. 

Without the hard leptons, the signal final states mostly consists of $\cancel{E}_{T}$.  The general searches of such invisible-inos at the LHC can be categorized into three sub-channels, indirect production from cascade decay of colored SUSY particles, direct pair production and weak boson fusion (WBF) produced pairs~\cite{Giudice:2010wb}. If the gluino/squarks are of several TeV, the production rate will be significantly suppressed. The direct pair production always requires an additional object to be produced associatively for the event trigger purpose and the final state is then mono-jet, mono-$Z$ or mono-photon plus $\cancel{E}_{T}$~\cite{Chen:1995yu,Feng:1999fu,Giudice:2010wb,Han:2013usa,Schwaller:2013baa,Cao:2009uw,Beltran:2010ww,Gori:2014oua,Goodman:2010ku,Rajaraman:2011wf,Fox:2011pm,ATLAS:2012ky,Chatrchyan:2012me,Baer:2014cua,Khachatryan:2014rra,Brooijmans:2014eja,Anandakrishnan:2014exa,Barducci:2015ffa,Mahbubani:2017gjh}.  However, this simple two-body final state lacks kinematic feature and it only demonstrates robust power when the production rate is huge, for instance, mono-jet plus stop pair in compressed stop-bino case. Due to the small signal rate for light Higgsino, the signal deviation is only at the comparable level as the systematic error as a few percent. The WBF production itself contains two forward-backward jets which can well serve as trigger  but suffers from small production rate of Higgsino pairs in comparison with the SM $2j+Z/W$ even with the WBF kinematic feature such as the tagging of two energetic forward-backward jet and central-jet veto, etc~\cite{Berlin:2015aba}. Recently, \cite{Egana-Ugrinovic:2018roi} discussed the Higgsino-like chargino search at colliders and found a Higgsino mass as light as 75 GeV still survived, assuming the mass splitting of chargino and lightest neutralino is less than 10 GeV.  Since the difficulty is that the $S/B$ ratio is within systematic error, the accumulate of LHC data won't improve the search \footnote{On the other hand, besides the direct limit at the colliders, the light Higgsino also receives constraints from dark matter direct detection experiments. In the limit of $\mu \ll M_{1},M_{2}$, the lightest neutralino $\tilde{\chi}^{0}_{1}$ can be identified as the dark matter candidate.~One the other hand, the pure Higgsino dark matter of 1~TeV produces the correct relic density.~Therefore, the light Higgsino only consists of a small portion of the dark matter and requires additional components such as axion.~As a result, a rescale of local abundance of Higgsino dark matter has to be taken into account into the calculation of direct detection bound. In the conventional natural SUSY framework, the light Higgsino dark matter is completely covered by the XENON1T experiment~\cite{Baer:2013vpa,Han:2013usa}.~However, this bound can be easily evaded by decoupling stop and $M_{1},M_{2}$.~For instance, if $\mu=120$~GeV, $M_{1}=M_{2}=2$~TeV, $m_{t_{L}}\sim m_{t_{R}}\sim 3$~TeV, $A_{t}=4.5$~TeV, the Higgs mass is 125.7~GeV, the rescaled Higgsino $\sigma_{SI}=0.05\times 10^{-46}~\text{cm}^{2}$ which is still one order of magnitude lower than the current XENON1T bound~\cite{Aprile:2017iyp}.  }.

To improve the $S/B$ ratio, we therefore study the search of such light Higgsinos at a future $e-p$ colliders.
Without color exchange between the colliding beams, $e-p$ colliders are the best places to study WBF processes with controlled backgrounds.~There are currently two proposed examples of such colliders. First one is an economic LHC upgrade $-$ the Large Hadron electron Collider (LHeC)~\cite{AbelleiraFernandez:2012cc}. 
By adding one electron beam of 60 -- 140~GeV to the current LHC {based on the 7~TeV proton beam} with a forward detector, LHeC is a future DIS facility proposal but recently there also exists discussion ``Higgs factory'' or precision machine based on high luminosity LHeC \cite{Han:2009pe,Biswal:2012mp,Li:2017kfk}. 
A second example is the Future Circular Collider in hadron-electron mode (FCC-eh)\cite{Kuze:2018dqd}, a future collider having 60~GeV electron beam colliding with a 50~TeV proton beam. 
The light Higgsino pairs can be produced at these $e-p$ colliders via WBF with a lower $\sqrt{\hat{s}}$ comparing with the LHC. But at the same time, the disappearing of QCD $2j+W/Z$ background may potentially enhance the signal-to-background ($S/B$). 
   
The paper is organized as following: {In the next section, we discuss the survival parameter space for light Higgsinos from direct searches. It is followed by a section on the phenomenology of this collider search, which includes the signal and background events simulation, some kinematic distributions of final states and selection cuts.~In section~\ref{result}, we give results of the signal-to-background ($S/B$) and significance ($Z$) varying with different $\mu$ values.~We then conclude in the final section. }

\section{\label{pheno}Phenomenology of light Higgsinos at the WBF machines}
\subsection{Signal and backgrounds}
\begin{figure}[H]
\centering
\subfloat[]{\includegraphics[scale=0.3]{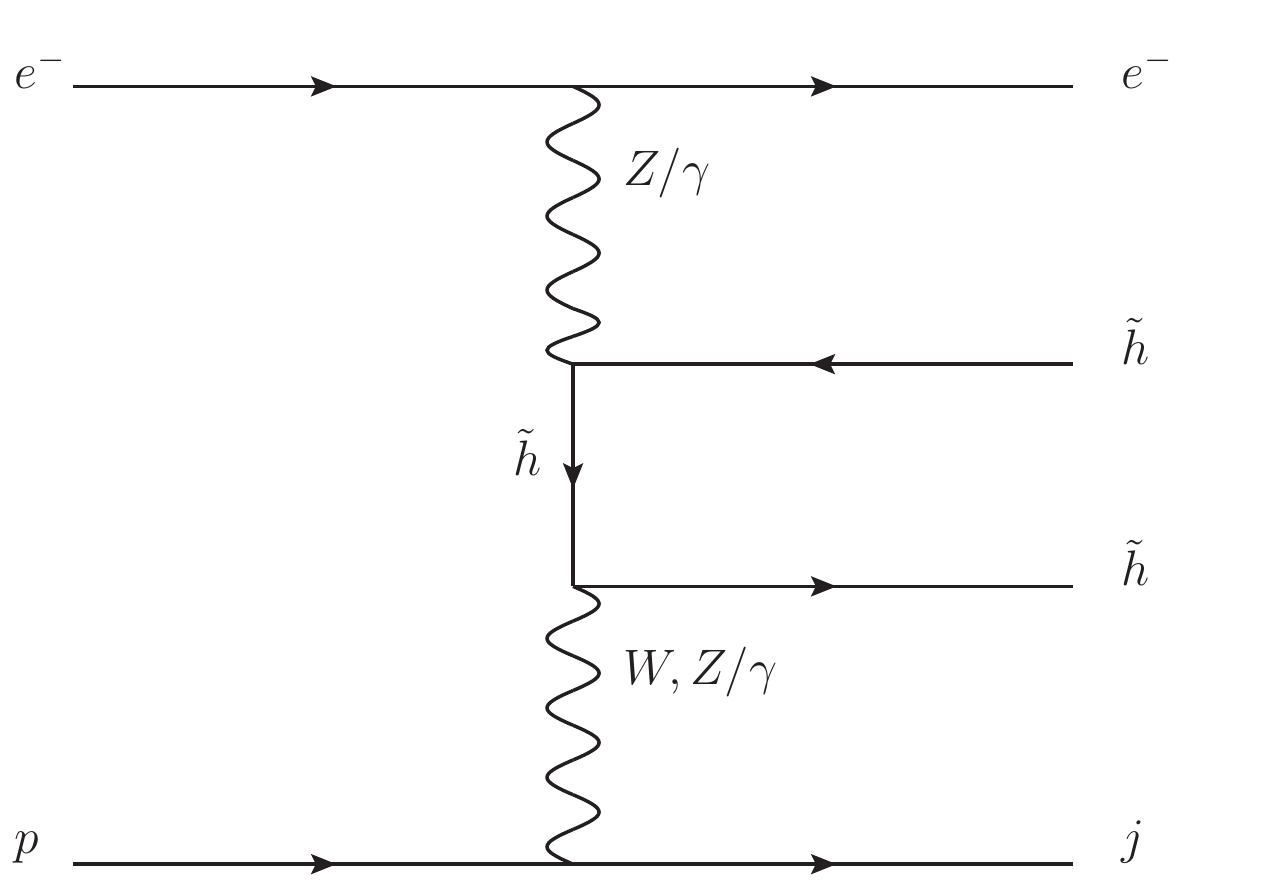}}
\subfloat[]{\includegraphics[scale=0.3]{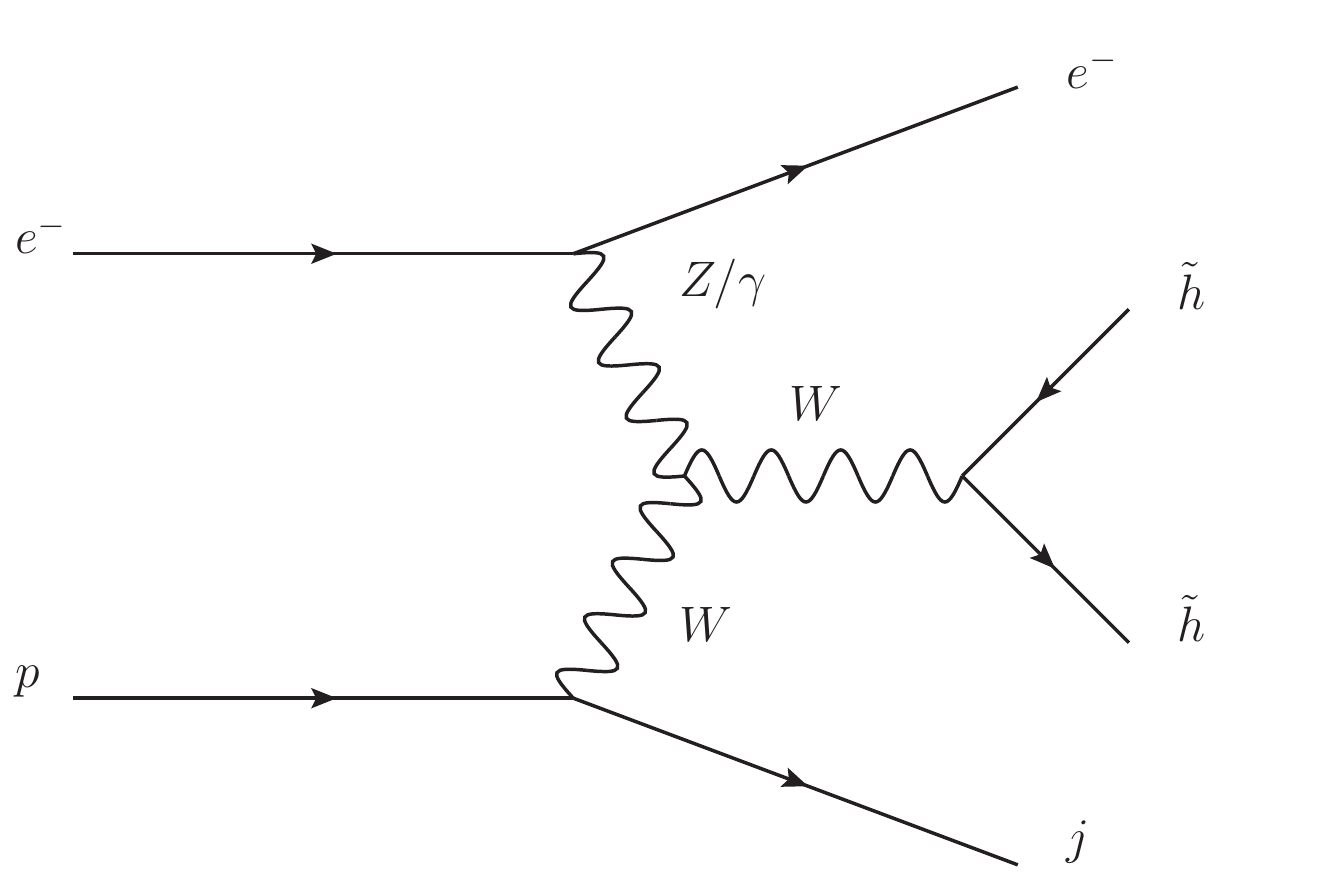}}
\subfloat[]{\includegraphics[scale=0.3]{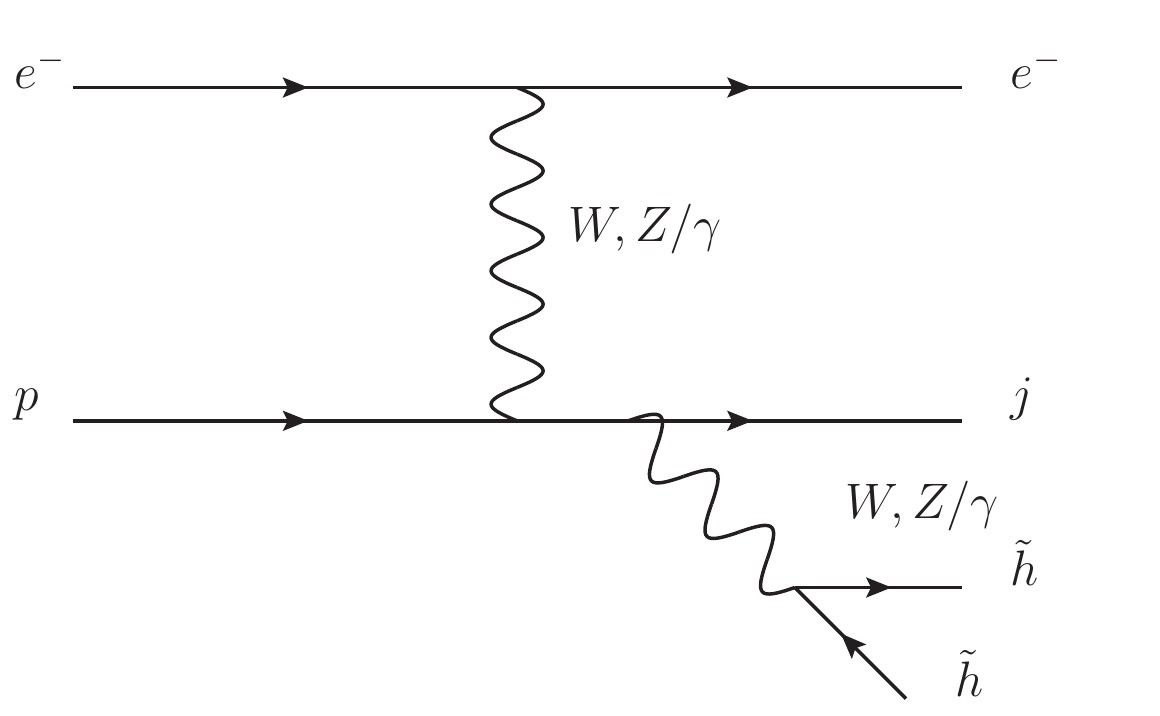}}
\subfloat[]{\includegraphics[scale=0.3]{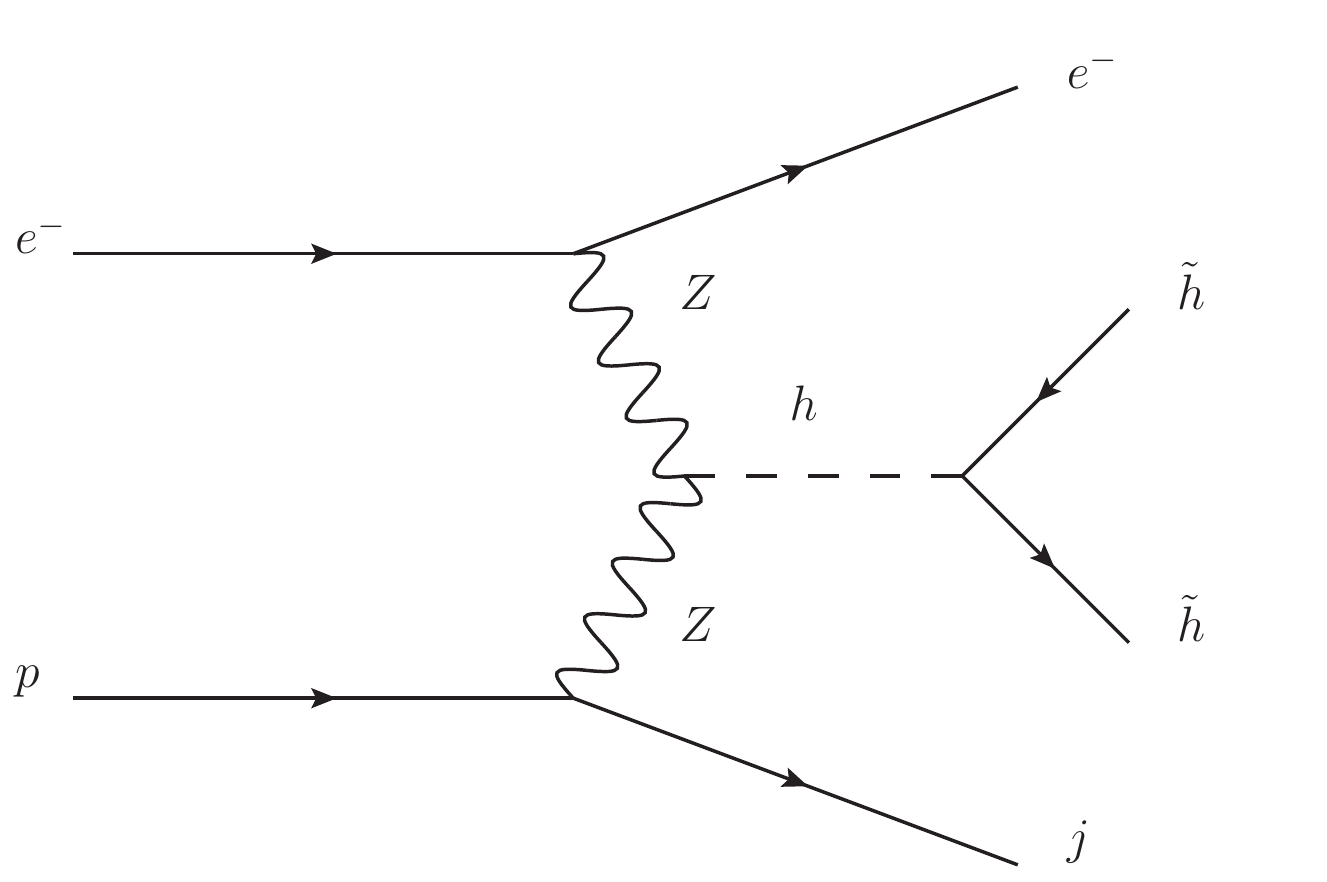}}
\caption{$h$ represents a Higgs boson, $\tilde{h}$ repesents a light Higgsino, i.e $\tilde{\chi}_{1}^{\pm}$ and $\tilde{\chi}_{1,2}^{0}$. }
\label{signaldiag}
\end{figure}

At the $e-p$ colliders, the SUSY particles are produced in pair via gauge interaction so the dominant contribution for Higgsino production is though weak boson fusion (WBF) processes. 
The light Higgsino contains three physical states $\tilde{\chi}^{\pm}_{1}$ and $\tilde{\chi}^{0}_{1,2}$ so  
the Higgsino production corresponds to the following channels: 
\bea
e^{-} p \rightarrow e^{-} j {\tilde{\chi}}_{1}^{\pm} {\tilde{\chi}}_{1}^{\mp}, e^{-} j {\tilde{\chi}}_{1}^{\pm} {\tilde{\chi}}_{1,2}^{0}, e^{-} j {\tilde{\chi}}_{1,2}^{0} {\tilde{\chi}}_{1,2}^{0} \nonumber~.
\eea
The contributions can be categorized into four topological diagrams shown in Fig.\ref{signaldiag}.

As argued in the previous section, the survivor scenario for the light Higgsino requires all the other SUSY particles to be generally heavy and decoupled. The mass spectrum shown in Table.\ref{hmass} is calculated by $\it Suspect$~\cite{Djouadi:2002ze} and $\it SUSY$-$\it HIT$~\cite{Djouadi:2006bz} where $M_{1}=M_{2}=2~\rm{TeV}$ and all the other SUSY particles are set to be 3 TeV. We found the neutral Higgs boson like-SM is essentially not affected by different $\mu$.~We focus on the range of $\mu \lesssim 200$~GeV. In this case, the mass splitting between ${\tilde{\chi}}_{1}^{\pm}$, ${\tilde{\chi}}_{1}^{0}$ and ${\tilde{\chi}}_{2}^{0}$ is small ($\Delta m \lesssim 5~\rm{GeV}$).~Consequently, the leptons from subsequent decay  ${\tilde{\chi}}_{1}^{\pm} \rightarrow W^{\pm*}{\tilde{\chi}}_{1}^{0}$, ${\tilde{\chi}}_{2}^{0} \rightarrow Z^{*}{\tilde{\chi}}_{1}^{0}$ are too soft to be detected. So all the ${\tilde{\chi}}_{1}^{\pm}$ and ${\tilde{\chi}}_{2}^{0}$ final states will be identified as  $\cancel{E}_{T}$ as the dark matter candidate ${\tilde{\chi}}_{1}^{0}$.
\begin{table}[H]
\renewcommand\arraystretch{0.7}
\begin{center}
\scalebox{0.9}{
\begin{tabular}{|c|c|c|c|c|c|c|}
\hline
{\diagbox{Higgs boson}{ parameter $\mu$}}&{$\mu=80~$GeV}&{$\mu=90~$GeV}&{$\mu=100~$GeV}&{$\mu=110~$GeV}&{$\mu=120~$GeV}&{$\mu=130~$GeV}\\\hline
$h$&{$125.52~$GeV}&{$125.52~$GeV}&{$125.51~$GeV}&{$125.50~$GeV}&{$125.49~$GeV}&{$125.48~$GeV}\\\hline
$H$&{$3000.09~$GeV}&{$3000.09~$GeV}&{$3000.09~$GeV}&{$3000.09~$GeV}&{$3000.09~$GeV}&{$3000.09~$GeV}\\\hline
$A$&{$3000.01~$GeV}&{$3000.01~$GeV}&{$3000.01~$GeV}&{$3000.01~$GeV}&{$3000.01~$GeV}&{$3000.01~$GeV}\\\hline
$H^{+}$&{$3001.45~$GeV}&{$3001.45~$GeV}&{$3001.44~$GeV}&{$3001.44~$GeV}&{$3001.43~$GeV}&{$3001.43~$GeV}\\\hline
\hline
{\diagbox{Higgs boson}{ parameter $\mu$}}&{$\mu=140~$GeV}&{$\mu=150~$GeV} &{$\mu=160~$GeV}&{$\mu=170~$GeV}&{$\mu=180~$GeV}&{$\mu=190~$GeV}\\\hline
$h$&{$125.47~$GeV}&{$125.47~$GeV}&{$125.46~$GeV}&{$125.46~$GeV}&{$125.45~$GeV}&{$125.45~$GeV}\\\hline
$H$&{$3000.09~$GeV}&{$3000.09~$GeV}&{$3000.09~$GeV}&{$3000.09~$GeV}&{$3000.09~$GeV}&{$3000.09~$GeV}\\\hline
$A$&{$3000.01~$GeV}&{$3000.01~$GeV}&{$3000.01~$GeV}&{$3000.01~$GeV}&{$3000.01~$GeV}&{$3000.01~$GeV}\\\hline
$H^{+}$&{$3001.43~$GeV}&{$3001.42~$GeV}&{$3001.42~$GeV}&{$3001.42~$GeV}&{$3001.41~$GeV}&{$3001.41~$GeV}\\\hline
\end{tabular}
}
\caption{We state that $M_{1}=M_{2}=2$~TeV, $m_{t_{L}}\sim m_{t_{R}}\sim 3$~TeV, $A_{t}=3.2$~TeV here. $h$ is the neutral Higgs boson like-SM.}
\label{hmass}
\end{center}
\end{table}

In our study, the Monte-Carlo events are generated by  {\it MadGraph5\_v2.5.5}~\cite{Alwall:2014hca} at parton level. We implement  {\it Pythia6.420} \cite{Sjostrand:2006za} and {\it Delphes3.3.0} \cite{deFavereau:2013fsa} for parton shower and detector simulations respectively. At parton level, we require following basic cuts 
\begin{itemize}
\item $p_{T}^{j}>20~\rm{GeV}$
\item $p_{T}^{\ell}>5~\rm{GeV}$
\item $|\eta_{\ell,j}|<5$
\item $\Delta{R}_{j \ell}>0.4$ and $\Delta{R}_{\ell \ell}>0.4$~.
\end{itemize}

\begin{figure}[H]
\centering
\subfloat[LHeC]{\includegraphics[width=0.402\textwidth]{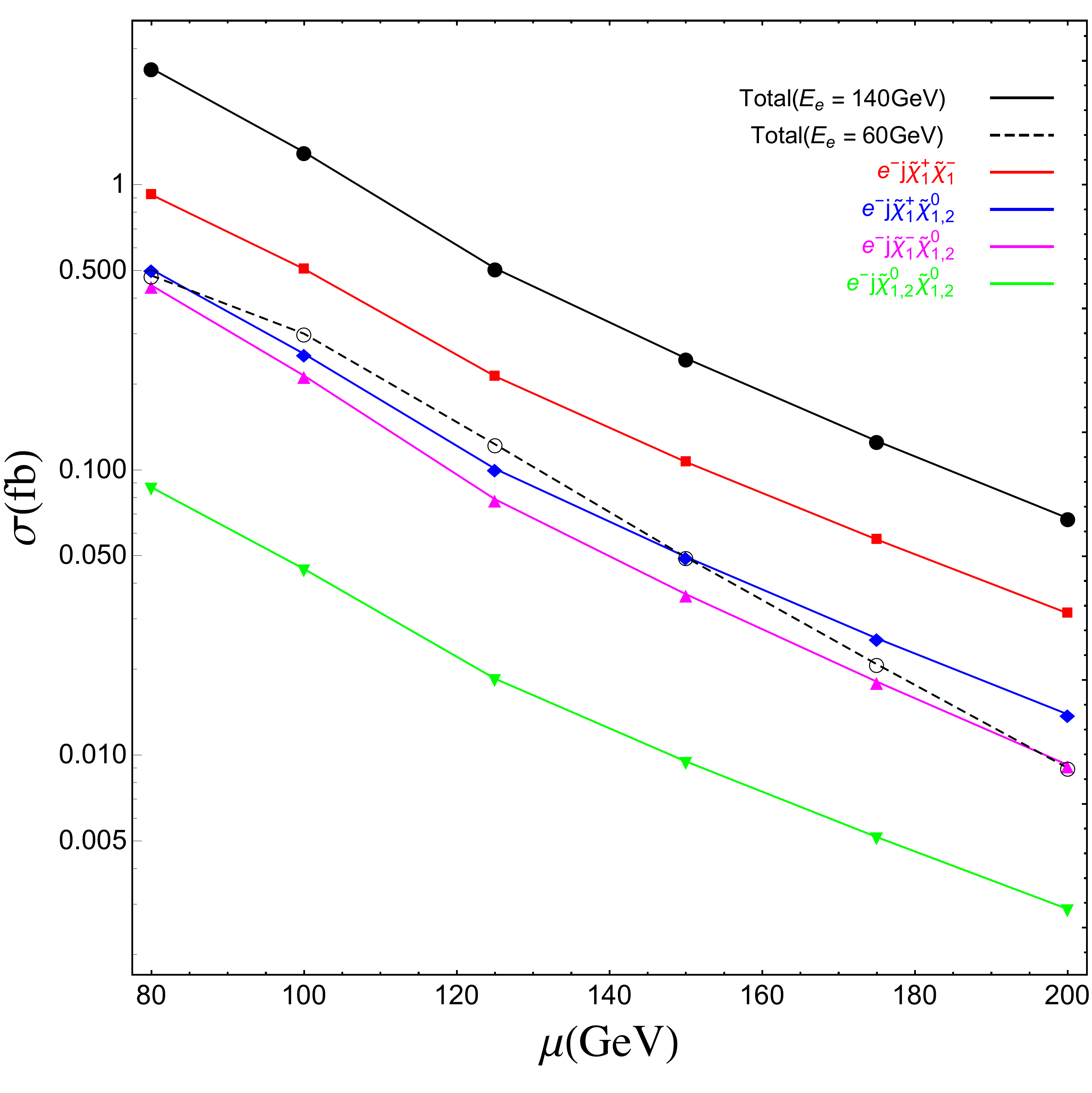}}
\subfloat[FCC-eh]{\includegraphics[width=0.40\textwidth]{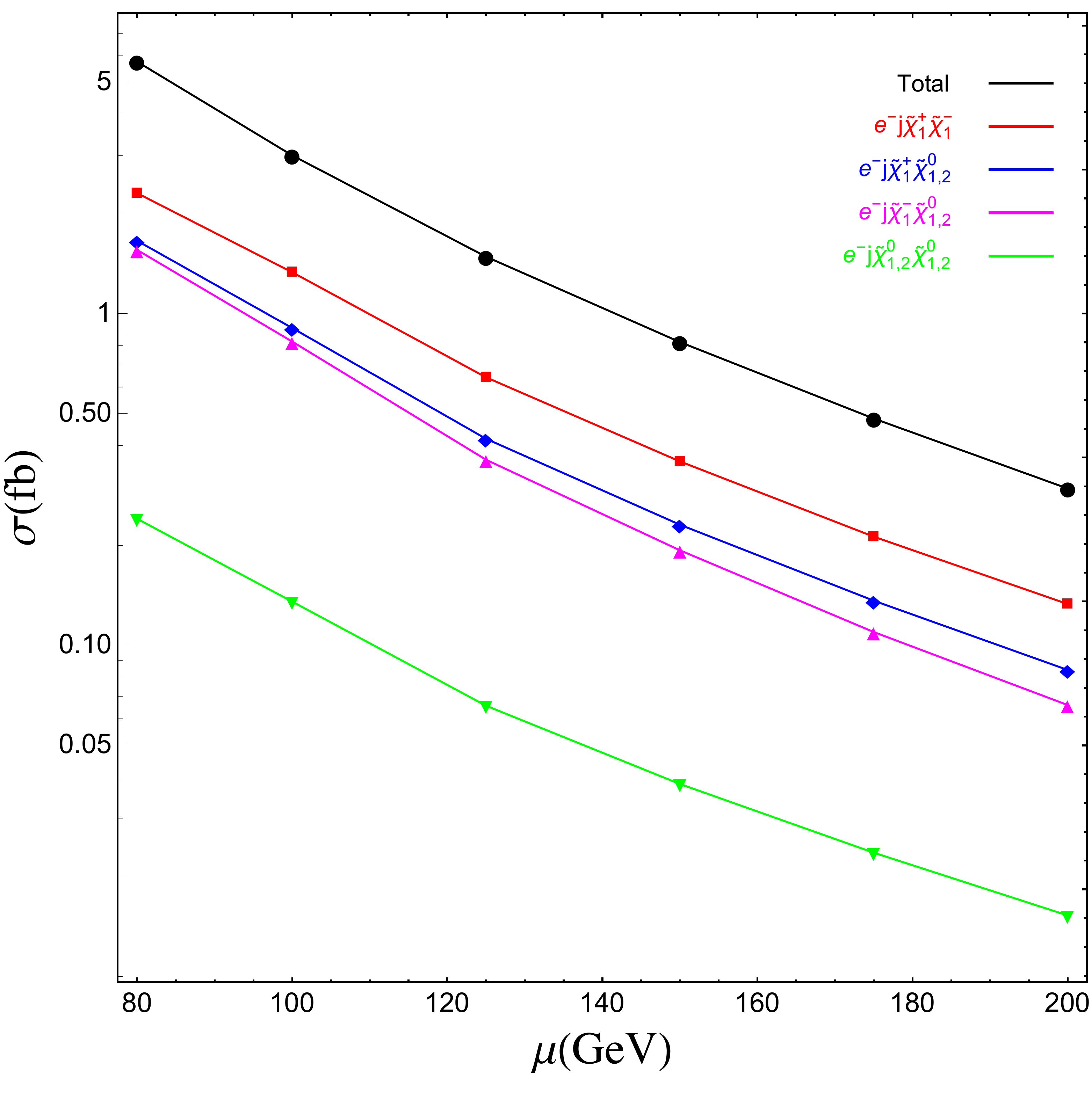}}
\caption{The partonic level cross sections of the signals varying with the Higgsino mass $\mu$ at LHeC (a) and FCC-eh (b). The combination of the cross section of all final states in the signal is defined as ``Total''. Where the solid lines represent the signal when $E_{e}=140~\rm{GeV}$, while the dashed line represents $E_{e}=60~\rm{GeV}$ in left panel. }
\label{cross}
\end{figure}

The production cross sections for Higgsino pairs as a function of~Higgsino mass $\mu$ is given in Fig.\ref{cross} for different colliders, $E_{e}$ and channels respectively. When $E_{e}=140(60)~\rm{GeV}$, the total cross section can be as large as $1.3(0.38)~{\text{fb}}$ for $\mu=100$~GeV. We also show a dependence on the cross section for electron $p_T > 5, 10, 15$ GeV in Table.\ref{pte} to agrue that the result would not break the perturbative description~\cite{Degrande:2016aje} even thought we set $p_{T}^{\ell}>5~\text{GeV}$ in basic cuts. With the enlarged centre-of-mass energy at the FCC-eh, the production rates are also significantly enhanced. 

\begin{table}[H]
\renewcommand\arraystretch{0.7}
\begin{center}
\begin{tabular}{|c|c|c|c|}
\hline
{\diagbox{beam energy}{ electron $p_{T}$}}&{$>5~$GeV}&{$>10~$GeV} & $>15~$GeV \\\hline
$140~\text{GeV}\times7000~\text{GeV}$ &  $1.30~$fb & $1.04~$fb & $0.89~$fb \\\hline
$60~\text{GeV}\times7000~\text{GeV}$ & $0.38~$fb &  $0.30~$fb & $0.25~$fb \\\hline
\end{tabular}
\caption{The cross section varying with the lower limit of electron $p_{T}$ with $E_{e}=140~\text{GeV}$ and $60~\text{GeV}$ respectively.}
\label{pte}
\end{center}
\end{table}
Since all the light Higgsino states are invisible in the detector, the final states consist of forward/backward jet and electron with $\cancel{E}_{T}$. The irreducible background can be classified into two categories: 
\bea
&&e^{-} p \rightarrow e^{-} j \nu_{e} \bar{\nu}_{e} \nonumber \\
&&e^{-} p \rightarrow e^{-} j \nu_{\mu,\tau} \bar{\nu}_{\mu,\tau} \nonumber
\eea 
where the invisible SM neutrinos makes the $\cancel{E}_{T}$. The backgrounds in the first category mainly come from $W$ bremsstrahlung process  ${e^{-} p} \rightarrow \nu_eW^{-}j$ with $W^- \rightarrow e^{-} \bar{\nu}_{e}$, and also partly from $Z$ bremsstrahlung process  ${e^{-} p}  \rightarrow e^{-}Zj$ with $Z \rightarrow \nu_{e} \bar{\nu}_{e}$ as well as the interference terms between them. While backgrounds in the second category is purely from ${e^{-} p}  \rightarrow e^{-}Zj$ production with $Z \rightarrow \nu_{\mu,\tau} \bar{\nu}_{\mu,\tau}$. The total cross sections of these two backgrounds are 243.6(115.5) $\text{fb}$ and 58.11(32.82) $\text{fb}$ when $E_{e}=140(60)~\rm{GeV}$ respectively, which are hundreds times larger than the Higgsinos production. Effective cuts are needed to reduce the huge backgrounds.

For the reducible background, the production of $\tau$ in the final state would be the largest one
\bea
&&e^{-} p \rightarrow e^{-} j \tau^{+} {\nu}_{\tau}, \nonumber \\
&&e^{-} p \rightarrow e^{-} j \tau^{-} \bar{\nu}_{\tau}. \nonumber
\eea 
The total cross sections are 330.2(163.8) $\text{fb}$ and 302.0(146.8) $\text{fb}$ when $E_{e}=140(60)~\rm{GeV}$ respectively. There are two main cases that these processes might fake the signal. (i) The $\tau$ fakes a hard jet in the detector, or (ii) the products from hadronic $\tau$ decays are too soft to be tagged contributing to the $\slashed{E}_{T}$. In fact the latter one is similar to the contribution of the irreducible background. Of course leptonic $\tau$ decay would contribute to our process as well, but taking into account of the existence of extra leptons in the final states, we can cut this background easily by vetoing the event with extra hard leptons. The other distinct reducible backgrounds are  
$e^{-} p \rightarrow j {\nu}_{e} \tau^{+} {\nu}_{\tau}$ and 
$e^{-} p \rightarrow j {\nu}_{e} \tau^{-} \bar{\nu}_{\tau}$.
They might mimic the signal since there is one electron from leptonic $\tau$ decays. Fortunately, we could suppress them to an insignificant order because of the totally different kinematic distribution of the final electron. According to the above analysis these backgrounds would not be considered in the following.

A left-handed polarization of electron beam may in principle enhance the signal production rate but the background will be enlarged at the same time\cite{Mondal:2015zba,Duarte:2018xst}. The polarization of electron beam does not significantly improve the result so we focus on the un-polarized cases.

\subsection{Selection cuts}

In order to suppress reducible backgrounds, we adopt following veto criteria at the LHeC \footnote{Extra leptons would also appear in the signal from chargino decays, but they are very soft because of compressed mass spectrum.}  : 
\begin{itemize}
\item[i.] We veto events containing any central jets $p_T^{{j_i}}> 3.0$ GeV, $|\eta_{j_{i}}|<2.0$ ($i\geq2; i \in \mathbb{N}$);
\item[ii.] A veto on events with extra hard electrons with $p_{T}^{e_{m}}>5.0~\text{GeV}$ ($m\geq2;m\in \mathbb{N}$) or muons with $p_{T}^{\mu_{k}}>5.0~\text{GeV}$ ($k\geq1;k \in \mathbb{N}$);
\item[iii.] Any events with a tagged $\tau$-jet is vetoed.  
\end{itemize}

Since the signal final state contains two massive invisible Higgsinos which presumably leads to large $\cancel{E}_{T}$, a cut on $\cancel{E}_{T}$ may help to distinguish the signal and backgrounds.

\begin{figure}[H]
\centering
\subfloat[$E_{e}=60~\rm{GeV}$]{\includegraphics[width=0.52\textwidth]{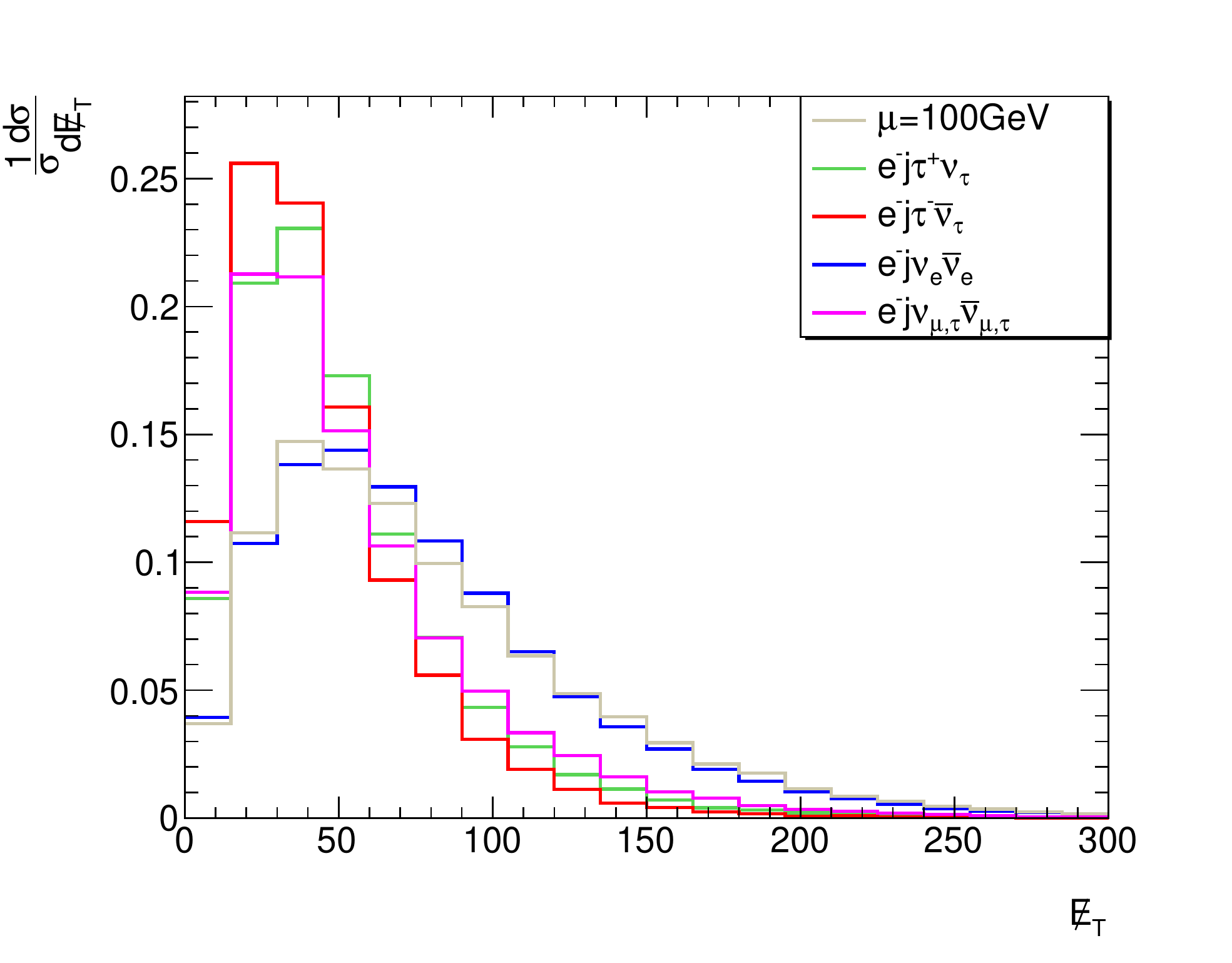}} 
\subfloat[$E_{e}=140~\rm{GeV}$]{\includegraphics[width=0.52\textwidth]{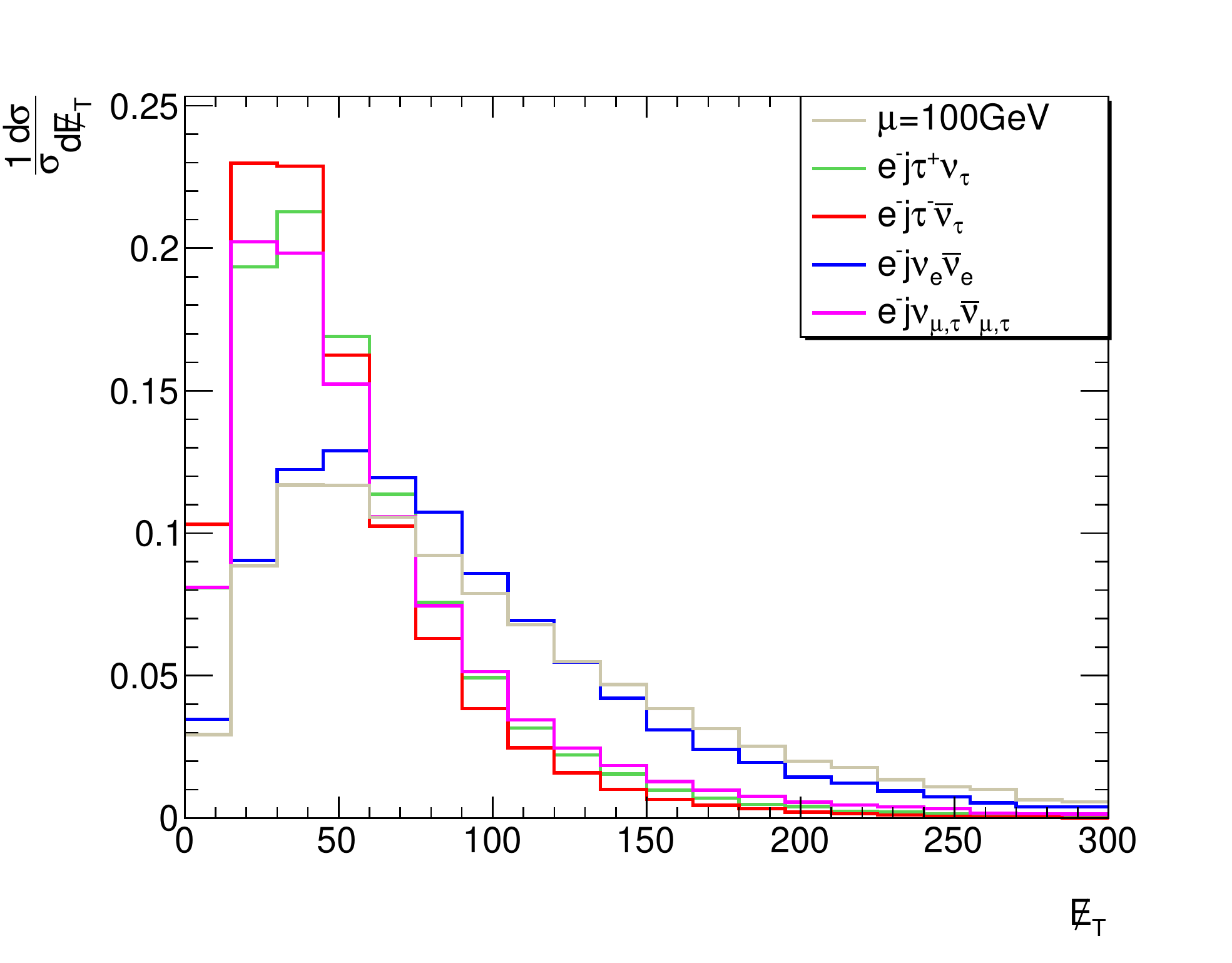}}
\caption{The normalised missing transverse energy $\slashed{E}_{T}$ distributions when $E=60~\rm{GeV}$ (left panel) and $E=140~\rm{GeV}$ (right panel) respectively.}
\label{met}
\end{figure}
In Fig.\ref{met}~we show the missing transverse energy distributions of the signal and backgrounds,~where the blue and magenta lines correspond to the above two irreducible backgrounds, the green and red lines correspond to the two irreducible backgrounds, while the grey line corresponds to the $\mu = 100$~GeV.
It is clearly shown that the $e^{-} j \nu_{\mu,\tau} \bar{\nu}_{\mu,\tau}$ and two $\tau$ backgrounds have a smaller missing transverse energy than the signal due to the small recoil mass.~While $e^{-} j \nu_{e} \bar{\nu}_{e}$ tend to have larger missing transverse momentum because the missing energy is not only from $W$ decay but also from  scattering with the initial electron. On the other hand, the missing transverse energy tend to be larger when the electron beam energy increases, as shown in Fig.\ref{met}.

Comparing the signal with the background events, the signal electron in WBF final state is forward, while in the leading SM background electrons are from $W^{-}$ decay. Therefore, one expects the leading electron has larger rapidity but less $p_{T}$ in the signal. For the background, leading electrons from $W^{-}\rightarrow e^{-} \bar{\nu}_{e}$ have larger $p_T$ as the Jacobian peak $m_{W}/2$ and are more in the central region.  

\begin{figure}[H]
\centering
\subfloat[The pseudo-rapidity $\eta_{e_{1}}$ distributions]{\includegraphics[width=0.52\textwidth]{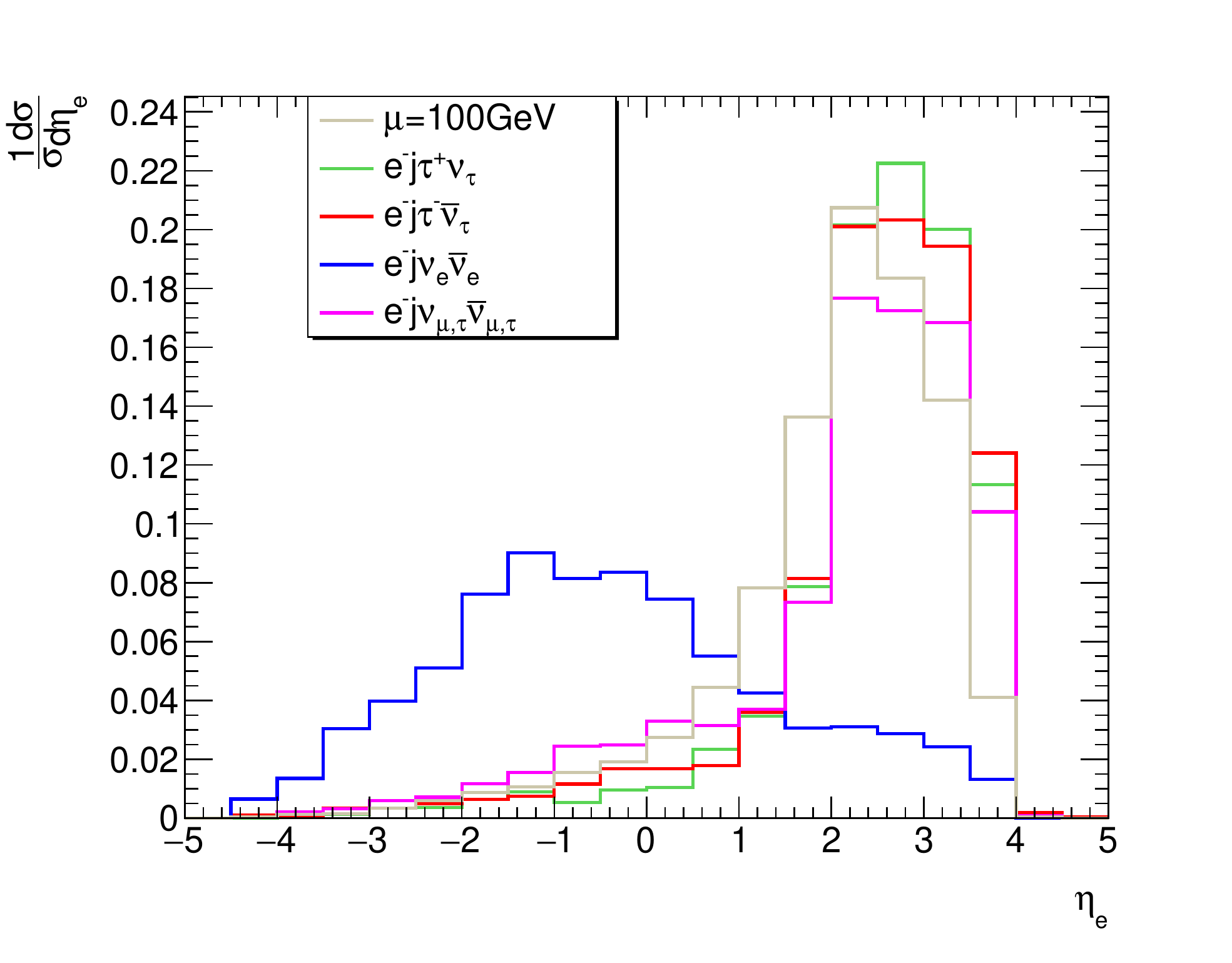}}
\subfloat[The the pseudo-rapidity $\eta_{j_{1}}$ distributions]{\includegraphics[width=0.52\textwidth]{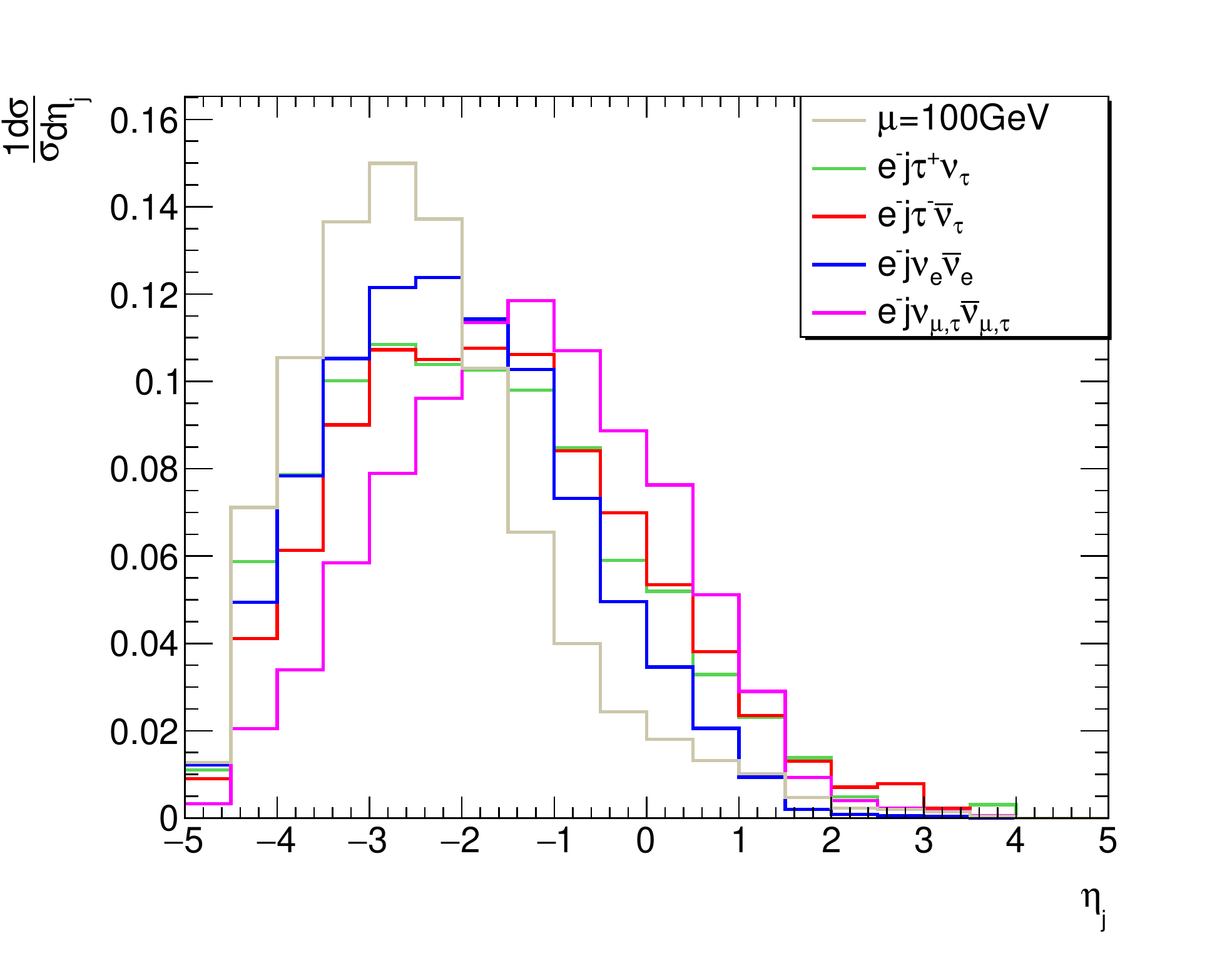}}\\
\subfloat[The invariant mass $m_{ej}$ distributions]{\includegraphics[width=0.52\textwidth]{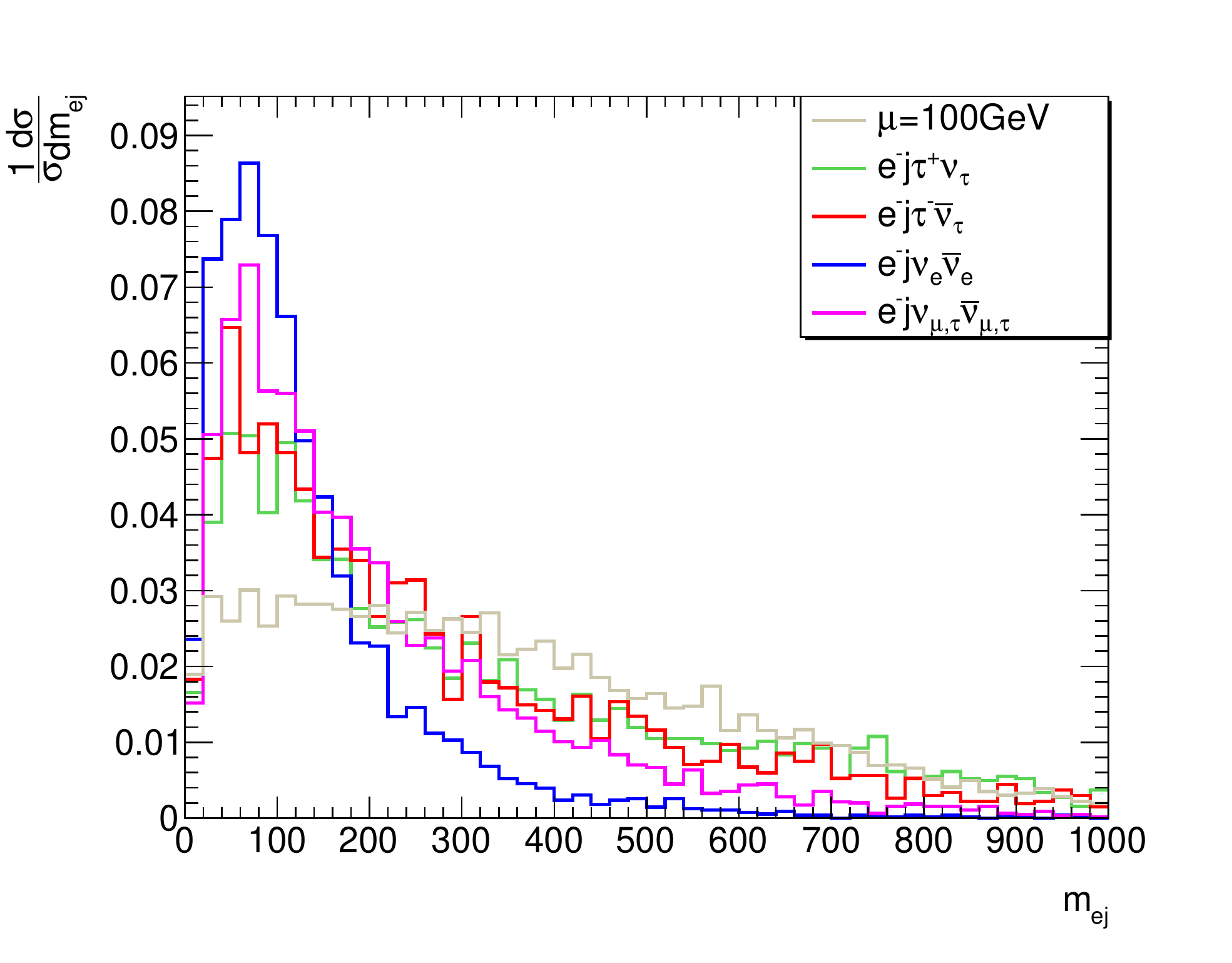}}
\subfloat[The inelasticity variable $y$ distributions]{\includegraphics[width=0.52\textwidth]{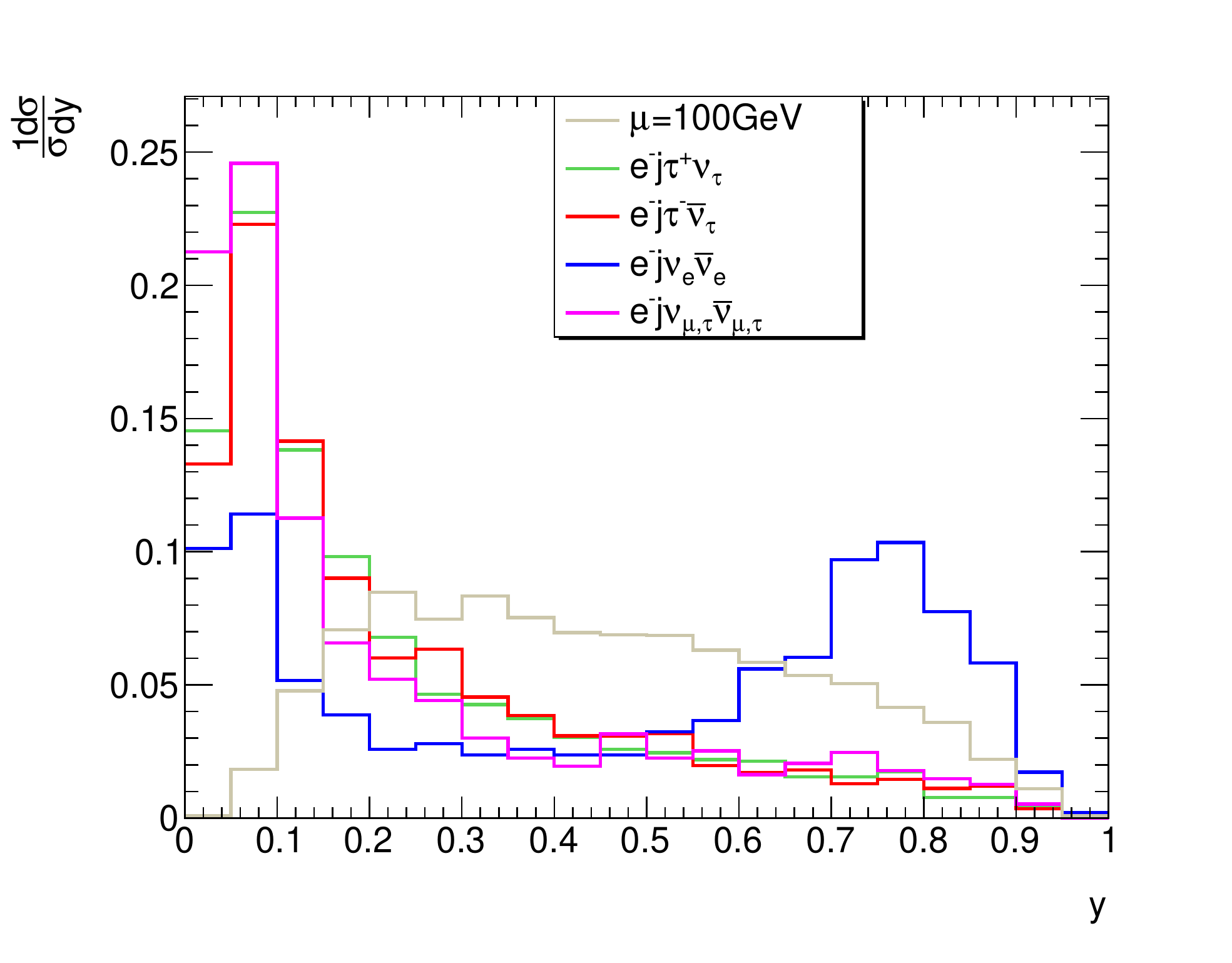}}
\caption{The normalised pseudo-rapidity $\eta_{e_{1}}$ (top left), $\eta_{j_{1}}$ (top right) distributions and the invariant mass $M(e_{1},j_{1})$ (bottom left) after veto criteria cuts i-ii when $E_{e}= 140$ GeV. The normalised inelasticity variable $y$ (bottom right) distributions after veto criteria cuts i-iii when $E_{e}= 140$ GeV.}
\label{mej-y}
\end{figure}

Another typical WBF cut is the invariant mass between the forward/backward jet/electron $M(e_{1},j_{1})$.  
The inelasticity variable $y$ introduced in \cite{Tang:2015uha}  is defined as 
\beq
y=\frac{k_{P} \cdot (k_{e}-p_{e})}{k_{P}\cdot k_{e}}
\eeq
where $k_{P}$ is the 4-momenta of the initial proton, $k_{e}$ is the 4-momenta of the initial electron, $p_{e}$ is the 4-momenta of the out-going electron.~This inelasticity variable is used to reflect the momentum transformation between the initial and final states.~In Fig.\ref{mej-y}, the pseudo-rapidity $\eta_{e_{1}}$, $\eta_{j_{1}}$, the invariant mass $M(e_{1},j_{1})$ and the inelasticity $y$ distributions are plotted respectively.  

Therefore, additional cuts are imposed as the following:
\begin{itemize}
\item[i.] Missing transverse energy cut: $\slashed{E}_{T}>30~\rm{GeV}$.
\item[ii.] Transverse momentum of the leading electron~$p_{T}^{e_{1}}<30~\rm{GeV}$.  
\item[iii.] Pseudorapidity of the leading electron and the forward leading jet in the final state~$\eta_{e_{1}}>1.0,\eta_{j_{1}}<-2.0$ .  
\item[iv.] The invariant mass of the leading electron and the forward leading jet in the final state $M(e_{1},j_{1}) >400~\rm{GeV}$. 
\item[v.] Inelasticity variable cut:  $y>0.3$ .
\end{itemize}

We presented the cut-flow for signal and background events in Table~\ref{cut-flow} for a $140$~{GeV} electron beam energy with an integrated luminosity of $1~\text{ab}^{-1}$. 
\begin{table}[H]
\renewcommand\arraystretch{0.7}
\centering
\begin{tabular}{|c||c|c|c|c|c|}
\hline
\hline
cut & $e^{-}j \tau^{+} {\nu}_{\tau}$ & $e^{-}j \tau^{-}\bar{\nu}_{\tau}$ & $e^{-}j \nu_{e}\bar{\nu}_{e}$ & $e^{-}j\nu_{\mu,\tau}\bar{\nu}_{\mu,\tau}$ & $\mu=100GeV$ \\

\hline
\hline
basic cuts & 330100 & 302000 & 243600 & 58110 & 1300 \\
\hline
detector efficiency  & 210340 & 189475 & 156055 & 35783 & 832.75 \\
\hline
veto criteria & 52446.3 & 43572.6 & 116056 & 24085.4 & 590.20 \\
\hline
$\slashed{E}_{T}>30~\rm{GeV}$ & 38073.7 & 28925.6 & 101849 & 16794 & 514.10 \\
\hline
$p_{T}^{e_{1}}<30~\rm{GeV}$ & 21482.9 & 16163 & 27034.7 & 7489.87 & 267.23 \\
\hline
$\eta_{j_{1}}<-2.0 \& \eta_{e_{1}}>1.0$ & 10219.9 & 7042.64 & 2260.61 & 2208.18 & 161.64 \\
\hline
$M(e_{1},j_{1})>400~\rm{GeV}$ & 5631.51 & 3412.60 & 570.02 & 728.70 & 80.02\\
\hline
$y>0.3$ & 805.44 & 567.76 & 253.34 & 58.11 & 48.54 \\

\hline
\hline
\end{tabular}
\caption{Cut-flow of the signal and background events for $\mu$= 100~GeV at 140~GeV electron beam energy LHeC with $\mathcal{L}= 1~\text{ab}^{-1}$. }
\label{cut-flow}
\end{table} 
After cuts, the total backgrounds are reduced by $\mathcal{O}(10^{-3})$ while the signal is still remaining around 4\%.~It is shown that after the parton shower and detector simulations, both of the signal and backgrounds have a $60\text{\%}-70$\% survival probability.~Veto criteria will suppress two reducible backgrounds by a factor of four and the requirements of $\eta_{e_{1}, j_{1}}$ dramatically reduce the $\nu_{e}\bar{\nu_{e}}$ background by approximate one order.~After the inelasticity variable cut, the remaining reducible background events is small and could be compared to the signal.~However, the $\tau$ background is still about 30 times larger than the signal, which could lead to a dramatic decline in the significance in particular for the 60~GeV beam case. In order to increase the production cross section of the signal, we put forward our study at the future higher energy $e-p$ collider, FCC-eh, to more clearly indicate the feature of searching the light Higgsino via VBF process.

As mentioned in the Introduction, FCC-eh collides a $60$~GeV electron beam with $50$~TeV proton beam of the planned Future Circular Collider (FCC), which is a typical deep inelastic facility with $\sqrt{s} \approx 3.5~\text{TeV}$.~Because of its higher energy and smaller QCD backgrounds, some precision physics and probing supersymmetric particles have been proposed recently\cite{Britzger:2017fuc,Schwanenberger:2018ilr,Das:2018vuk}. According to the above analysis, we focus on our process at the FCC-eh to compensate for the drawback of the small production cross section of the signal at the LHeC. The total cross section of the different backgrounds and signal are shown in Table~\ref{xsection}.

\begin{table}[H]
\renewcommand\arraystretch{0.8}
\centering
\begin{tabular}{|c||c|c|c|c|c|}
\hline
\hline
ep collider & $e^{-}j \tau^{+} {\nu}_{\tau}$ & $e^{-}j \tau^{-}\bar{\nu}_{\tau}$ & $e^{-}j \nu_{e}\bar{\nu}_{e}$ & $e^{-}j\nu_{\mu,\tau}\bar{\nu}_{\mu,\tau}$ & $\mu=100GeV$ \\
\hline
\hline
LHeC with $E_{e}=60$~GeV & 163.8 & 146.8 & 115.5 & 32.82 & 0.38 \\
\hline
LHeC with $E_{e}=140$~GeV & 330.2 & 302.0 & 243.6 & 58.11 & 1.3 \\
\hline
FCC-eh & 546.5 & 567.0 & 446.6 & 100.7 & 3.0 \\
\hline
\hline
\end{tabular}
\caption{The production cross section of all backgrounds and the signal with $\mu=100~\rm{GeV}$ at different $e-p$ colliders setup respectively.}
\label{xsection}
\end{table} 

Since FCC-eh has a higher proton beam energy than LHeC, we need to modify the above veto criteria and kinematic cuts to adjust to the progressive detector simulation. The modified values for the cuts and cut-flow are presented in Table \ref{fcc-cuts} and \ref{fcc-cut-flow} respectively. It is clearly shown that we concentrate on more forward region of the final jets as well as larger missing transverse energy because of $50~\text{TeV}$ proton beam. $\slashed{E}_{T}$ and $p_{T}^{e_{1}}$ cuts are effective for all backgrounds reduced by one order. The last inelasticity variable is still a fine complementary cut without severe loss of signal. We can find the rest signal events in Table \ref{fcc-cut-flow} are approximately threefold than before at the LHeC with 140~GeV electron beam, while the increase in backgrounds is less than three times. We expect the significance has an obvious improvement at the FCC-eh.

\begin{table}[H]
\renewcommand\arraystretch{0.8}
\centering
\begin{tabular}{|c||c|}
\hline
\hline
basic cuts & the same \\
\hline
\hline
central jets veto &  $p_T^{{j_i}}> 3.0$ GeV and $ |\eta_{j_{i}}|<3.0$ ($i\geq2;i \in \mathbb{N}$) \\
\hline
hard extra leptons($e, \mu$) veto & $p_{T}^{e_{m}}>5.0~\text{GeV}$, $p_{T}^{\mu_{k}}>5.0~\text{GeV}$ ($m\geq2,k\geq1;m,k \in \mathbb{N}$) \\
\hline
$\tau\text{-jet}$ veto & vetoing any events with $\tau\text{-jet}$ \\
\hline
\hline
missing transverse energy cut & $\slashed{E}_{T}>70~\rm{GeV}$ \\
\hline
transverse momentum cut & $p_{T}^{e_{1}}<25~\rm{GeV}$ \\
\hline
pseudorapidity cuts & $\eta_{e_{1}}>0.0,\eta_{j_{1}}<-3.0$ \\
\hline
invariant mass cut & $M(e_{1},j_{1}) >400~\rm{GeV}$ \\
\hline
nelasticity variable cut & $y>0.15$ \\
\hline
\hline
\end{tabular}
\caption{The The modified veto criteria and kinematic cut for the FCC-eh.}
\label{fcc-cuts}
\end{table} 

\begin{table}[H]
\renewcommand\arraystretch{0.7}
\centering
\begin{tabular}{|c||c|c|c|c|c|}
\hline
\hline
cut & $e^{-}j \tau^{+} {\nu}_{\tau}$ & $e^{-}j \tau^{-}\bar{\nu}_{\tau}$ & $e^{-}j \nu_{e}\bar{\nu}_{e}$ & $e^{-}j\nu_{\mu,\tau}\bar{\nu}_{\mu,\tau}$ & $\mu=100GeV$ \\

\hline
\hline
basic cuts & 546500 & 567000 & 446600 & 100700 & 3000 \\
\hline
detector efficiency  & 346251 & 356121 & 285270 & 62017.1 & 1926.54 \\
\hline
veto criteria & 41329 & 36968 & 169842 & 30955.2 & 1069.08 \\
\hline
$\slashed{E}_{T}>70~\rm{GeV}$ & 13400.2 & 9888.48 & 103549 & 9810.19 & 737.22 \\
\hline
$p_{T}^{e_{1}}<25~\rm{GeV}$ & 4842 & 3742.2 & 16292 & 2469.16 & 283.44 \\
\hline
$\eta_{j_{1}}<-3.0 \& \eta_{e_{1}}>0.0$ & 3432.1 & 2381.4 & 946.66 & 846.01 & 203.88 \\
\hline
$M(e_{1},j_{1})>400~\rm{GeV}$ & 3169.7 & 2177.3 & 625.24 & 797.54 & 178.02\\
\hline
$y>0.3$ & 1103.93 & 635.04 & 303.69 & 130.91 & 138.1 \\
\hline
\hline
\end{tabular}
\caption{Cut flow of the signal and background events for $\mu$= 100~GeV at the FCC-eh with $\mathcal{L}= 1~\text{ab}^{-1}$. }
\label{fcc-cut-flow}
\end{table}

\section{\label{result}Results}
We calculate the signal significance $Z$ after full detector simualtion through the formula:
\bea
Z=\frac{S}{\sqrt{S+B}} \nonumber
\eea
where $S$ repsesents the number of signal events, $B=\Sigma_{i}B_{i}$ denotes the overall background ($i=e^{-}j \tau^{+} {\nu}_{\tau}, e^{-}j \tau^{-}\bar{\nu}_{\tau}, e^{-} j \nu_{e} \bar{\nu}_{e}$, $e^{-} j \nu_{\mu.\tau} \bar{\nu}_{\mu,\tau}$). The signal-to-background ($S/B$) and significance ($Z$) dependent on different Higgsino masses and integrated luminosities are shown in the following Table~\ref{significance1}. When $\mu=80~\rm{GeV}$, the signal-to-background is about 5.1\%.~Even though $\mu$ up to $100~\text{GeV}$ the $S/B$ is still 2.7\%, and we expect it would not be overshadowed by uncertainties.~At $\mathcal{L}= 3 ~\text{ab}^{-1}$, the Higgsino mass could be probed over $2\sigma$ significance for $ \mu \lesssim 95~\text{GeV}$.~Note the above result are based on 140~GeV electron beam at the LHeC, while for 60~GeV beam energy we don't show the result since the $\tau$ backgrounds overshadow the tiny signal so that the sensitivity can not reach to $2\sigma$ after detector simulation. We also present the results of the FCC-eh compared to the LHeC. It shows that the higher energy would improve the significance dramatically. However, more comprehensive and complete analysis need to be performed with realistic detector.
\begin{table}[H]
\renewcommand\arraystretch{0.8}
\centering
\begin{tabular}{|c|c||c|c|c|}
\hline
\hline
\multicolumn{2}{|c||}{~} & $\mu=80~\rm{GeV}$ & $\mu=90~\rm{GeV}$ & $\mu=100~\rm{GeV}$ \\
\hline
\hline
\multirow{3}{*}{LHeC (140~GeV)} & $S/B$ & {5.1}\% & {3.5}\% & {2.7}\% \\
\cline{2-5}
 & $Z$ with $\mathcal{L}= 1 ~\text{ab}^{-1}$ & ${2.1}\sigma$ & ${1.5}\sigma$ & ${1.1}\sigma$ \\
\cline{2-5}
 & $Z$ with $\mathcal{L}= 3 ~\text{ab}^{-1}$ & ${3.6}\sigma$ & ${2.4}\sigma$ & ${1.8}\sigma$ \\
\hline
\hline
\multirow{3}{*}{FCC-eh} & $S/B$ & {10.3}\% & {8.3}\% & {6.6}\% \\
\cline{2-5}
& $Z$ with $\mathcal{L}= 1 ~\text{ab}^{-1}$ & ${4.5}\sigma$ & ${3.7}\sigma$ & ${3.0}\sigma$ \\
\cline{2-5}
& $Z$ with $\mathcal{L}= 3 ~\text{ab}^{-1}$ & ${7.9}\sigma$ & ${6.4}\sigma$ & ${5.3}\sigma$ \\
\hline
\hline
\end{tabular}
\caption{The signal-to-background and significance of Higgsino with $\mu=80, 90 $ and $100~\rm{GeV}$ with $\mathcal{L}= 1~\text{ab}^{-1}$ and $3~\text{ab}^{-1}$ at the LHeC with $E_{e}=140~\rm{GeV}$ and FCC-eh respectively.}
\label{significance1}
\end{table}

\begin{figure}[H]
\centering
\includegraphics[width=0.54\textwidth]{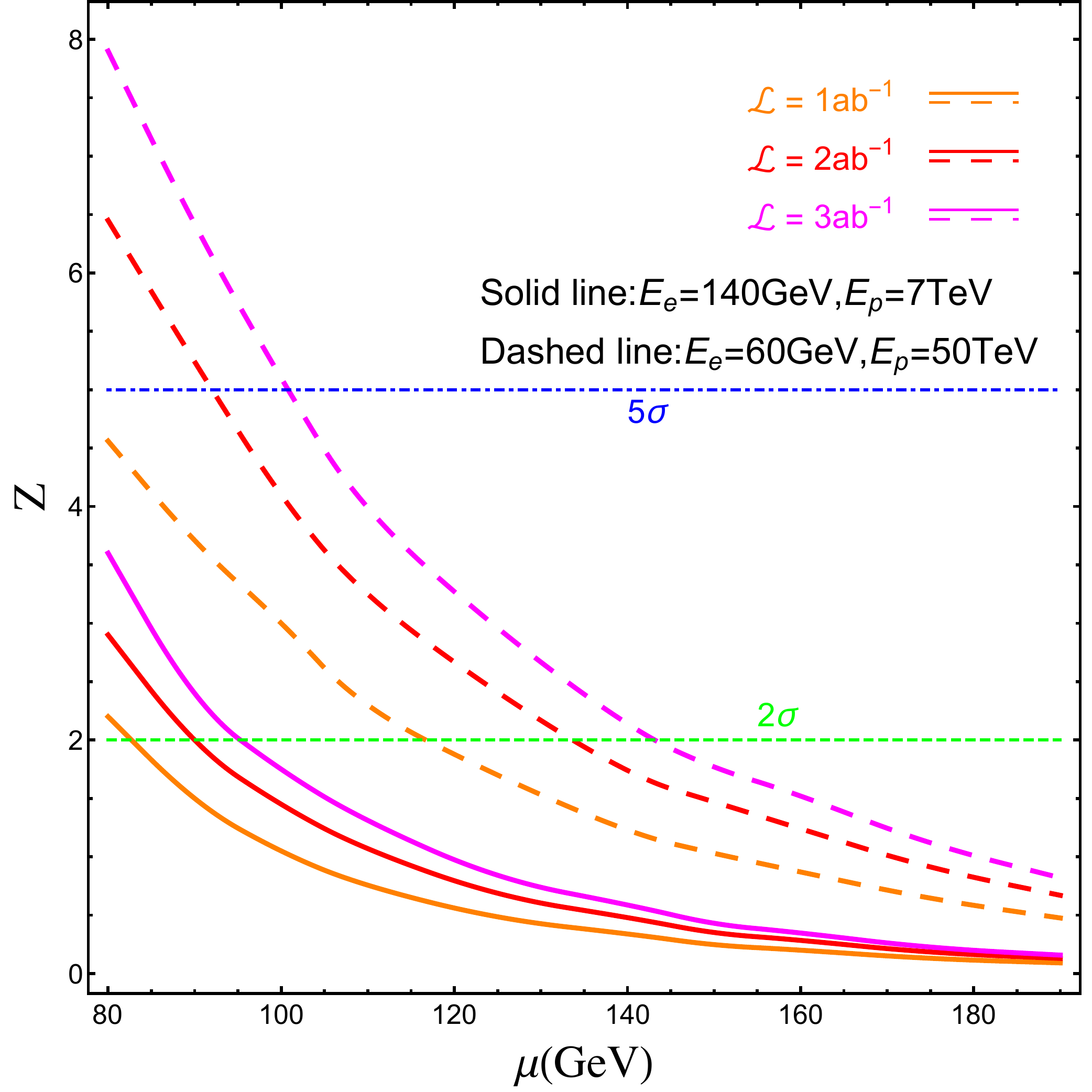}
\caption{The significance $Z$ varying with the Higgsino mass $\mu$ at the LHeC with $E_{e}=140~\rm{GeV}$ and FCC-eh respectively.}
\label{significance2}
\end{figure}

In Fig.\ref{significance2}, we show the dependence of the signal significance $Z$ on the Higgsino mass $\mu$ at the LHeC with $E_{e}=140~\rm{GeV}$ and FCC-eh respectively. With an increase of $\mu$ the significance drops rapidly due to the reduction in the signal cross sections. At the $140~\text{GeV}$ LHeC, $\mu=85(95)~\rm{GeV}$ with $\mathcal{L}= 1(3) ~\text{ab}^{-1}$ is the threshold for $2\sigma$ significance.~One can see the signal cross section is too small to have sensitivity to probe greater Higgsino mass even when we update an integrated luminosity to $3~\text{ab}^{-1}$. On the other hand, for FCC-eh we can easily reach to $2\sigma$ significance nearby $\mu=115~\rm{GeV}$ with $\mathcal{L}= 1 ~\text{ab}^{-1}$ and $\mu=145~\rm{GeV}$ with $\mathcal{L}= 3 ~\text{ab}^{-1}$ respectively. Higher electron beam energy and luminosity at the future ep collider would potentially probe heavier Higgsinos.

\section{Conclusion}
Searching for the existence the Higgsinos is crucial to test the criteria of the naturalness in supersymmetry.  It is proved to be difficult to search it at hadron colliders, due to the small signal rates while large systematic uncertainties. On the other hand, the electron positron collider may have better control of systematic uncertainties, however, the capabilities are limited by the central energy. In this paper, we discuss to look for Higgsinos at the future Large Hadron Electron Collider (LHeC), {which is a deep inelastic scattering (DIS) facility designed for a high precision measurement of patron distribution functions (PDFs) and forward physics detection.} Through our simulation, we find a mass of Higgsino around 85(95) GeV can be reached at 2$\sigma$ level with a luminosity  $1(3) ~\text{ab}^{-1}$, while at the FCC-eh we can reach to $2\sigma$ significance nearby $\mu=115(145)~\rm{GeV}$ with $\mathcal{L}= 1(3) ~\text{ab}^{-1}$. Heavier Higgsinos can be approached by increasing the luminosity and beam energy.

\section*{Acknowledgments}
{We thank LHeC Higgs\&Top group for sharing Delphes card with LHeC and FCC-eh setup.}~This work is supported in part by the National Science Foundation of China (11875232,~11422544); the World Premier International Research Center Initiative (WPI), MEXT, Japan and by the Zhejiang University K.P.~Chaos High Technology Development Foundation. CH would like to thank Zhejiang University for its hospitality when this work was initiated.

\bibliographystyle{utphysmcite}
\bibliography{Higgsino_v2}

\end{document}